\documentclass[preprint,aps]{revtex4}
\usepackage{amsmath}
\usepackage{dcolumn}
\usepackage{verbatim}
\usepackage{fancyvrb}
\usepackage{fancyhdr}

\begin{document}

\title{Wave functions in the $f_{7/2}$ shell, for educational purposes and
ideas}

\author{A.~Escuderos, L.~Zamick}
\affiliation{Department of Physics and Astronomy, Rutgers University,
Piscataway, New Jersey USA 08854}
\author{B.~F.~Bayman}
\affiliation{School of Physics and Astronomy, University of Minnesota,
Minneapolis, Minnesota 55455}

\date{\today}

\begin{abstract}
We use the spectrum of $^{42}$Sc as input to calculate wave functions and 
energy levels of all nuclei in the single-$j$-shell model with $j$ equal to 
$f_{7/2}$. The value of these results for educational purposes is pointed out. 
Also new insights emerge. For example, it is noted that for the $Z=4$, $N=4$ 
nucleus $^{48}$Cr, a good quantum number comes out: $(-1)^{(v_P+v_N)/2}$, where
$v_P$ and $v_N$ are the seniority quantum numbers of the protons and neutrons, 
respectively.
\end{abstract}

\maketitle

\section{Introduction}

In this work, we present energy levels and wave functions for nuclei in the 
$f_{7/2}$ shell. There was a previous work by McCullen, Bayman and 
Zamick~\cite{bmz63,mbz64} which was followed by a technical report that also
contained energy levels and wave functions. In this compilation, as in the
previous one, we take the two-body matrix elements from experiment. However, in
1964 some of the $T=0$ matrix elements were not known and guesses had to be 
made, some of which were incorrect; in this work, we use the correct spectrum. 
In the previous work, only the scandium and titanium nuclei, plus their cross
conjugates and mirrors, were included; in the present compilation, we consider
all the nuclei. For brevity, we do not include tables of wave functions for the
cross conjugate or mirror nuclei, but we provide formulas that show how these 
can be obtained from their partners.

At present, one can perform complete $fp$ shell calculations, so what is the
justification for calculations in such a truncated space? There is no question
but that one needs a complete $fp$ space to get quantitative agreement with 
experiment, and even that is not always sufficient---intruder states can play
an important role.

First of all, some properties do come out fairly well even in this small space.
More importantly, we feel that just looking at the wave functions and 
performing simple calculations with them is a valuable educational experience. 
We provide examples of points of interest. Beyond that, some interesting 
symmetries arise that suggest new ideas. The large shell model programs do not 
provide wave functions. We feel that looking at these wave functions can lead
to insights and surprises and indeed there will be some that the present 
authors have missed.

\section{Coefficients of fractional parentage}

In Tables~\ref{tab:cfp23} and~\ref{tab:cfp34}, we list the one-particle 
coefficients of fractional parentage (cfp's) used in this compilation. For a 
system of $n$ identical particles in a single $j$ shell with total angular 
momentum $I$, we have
\begin{equation}
\Psi (j^n v \alpha I M) = \sum_{v_1 \alpha_1} [j^{n-1} v_1 \alpha_1 J_1;j
|\} j^n v \alpha I] \left[ \Psi(j^{n-1} v_1 \alpha_1 J_1) \Psi(j) \right]^J_M.
\end{equation}
In the above, $\Psi$ is a totally antisymmetric wave function. The cfp's are
computed by the same method as was used by Bayman and Lande~\cite{bl66}. We 
refer the reader to this work for details. It should, however, be pointed out
that a different diagonalization routine is now being used to diagonalize the
relevant matrices. Hence, some of the sets of cfp's here can have an overall
sign difference from the original Bayman--Lande ones.

\begin{table}[ht]
\caption{$[7/2^2 K;7/2 |\}7/2^3 I]$.} \label{tab:cfp23}
\begin{tabular*}{.7\textwidth}{@{\extracolsep{\fill}}crrrr}
\toprule
 & \multicolumn{4}{c}{$K$} \\ \cline{2-5}
 & \multicolumn{1}{c}{0} & \multicolumn{1}{c}{2} & \multicolumn{1}{c}{4} 
 & \multicolumn{1}{c}{6} \\
$I$ \\
\colrule
$3/2$ & 0.000000 & $-0.462910$ & 0.886405 & 0.000000 \\
$5/2$ & 0.000000 & 0.781736 & 0.246183 & $-0.572960$ \\
$7/2$ & 0.500000 & $-0.372678$ & $-0.500000$ & $-0.600925$ \\
$9/2$ & 0.000000 & $-0.321208$ & 0.805823 & $-0.497468$ \\
$11/2$ & 0.000000 & $-0.527046$ & 0.443813 & 0.724743 \\
$15/2$ & 0.000000 & 0.000000 & $-0.476731$ & 0.879049 \\
\botrule
\end{tabular*}
\end{table}

\begin{table}[ht]
\caption{$[7/2^3 K;7/2 |\}7/2^4 v I]$} \label{tab:cfp34}
\begin{tabular*}{.9\textwidth}{@{\extracolsep{\fill}}ccrrrrrr}
\toprule
 & & \multicolumn{6}{c}{$K$} \\ \cline{3-8}
 & & \multicolumn{1}{c}{$3/2$} & \multicolumn{1}{c}{$5/2$} & \multicolumn{1}{c}
{$7/2$} & \multicolumn{1}{c}{$9/2$} & \multicolumn{1}{c}{$11/2$} & 
\multicolumn{1}{c}{$15/2$} \\ 
$I$ & $v$ \\ \colrule
0 & 0 & 0.000000 & 0.000000 & 1.000000 & 0.000000 & 0.000000 & 0.000000 \\
2 & 2 & 0.253546 & 0.524404 & $-0.577350$ & $-0.278174$ & 0.500000 & 0.000000\\
2 & 4 & 0.560612 & $-0.158114$ & 0.000000 & 0.754854 & 0.301511 & 0.000000 \\
4 & 2 & $-0.361873$ & 0.123091 & $-0.577350$ & 0.520157 & $-0.313823$ & 
 0.389249 \\
4 & 4 & $-0.175933$ & 0.658281 & 0.000000 & $-0.128388$ & 0.645497 & 0.320256\\
5 & 4 & $-0.387298$ & $-0.333712$ & 0.000000 & 0.553912 & 0.459390 & 
 $-0.469871$ \\
6 & 2 & 0.000000 & 0.238366 & 0.577350 & 0.267183 & 0.426401 & 0.597196 \\
8 & 4 & 0.000000 & 0.000000 & 0.000000 & 0.373979 & $-0.483494$ & 0.791438 \\
\botrule
\end{tabular*}
\end{table}

\section{The interaction}

We take the matrix elements from experiment, i.e., from the spectrum of 
$^{42}$Sc. We make the association
\begin{equation}
\langle (f^2_{7/2})^J | V | (f^2_{7/2})^J \rangle = E^*(J),
\end{equation}
where $E^*(J)$ is the {\bf experimental} excitation energy in $^{42}$Sc of the 
lowest state of angular momentum $J$.

The values of $E^*(J)$ in MeV are
\begin{center}
\begin{tabular*}{.5\textwidth}{@{\extracolsep{\fill}}rrcrr}
\toprule
\multicolumn{2}{c}{$T=1$} & & \multicolumn{2}{c}{$T=0$} \\ 
\cline{1-2} \cline{4-5}
$J=0$ & 0.0000 & & $J=1$ & 0.6111 \\
$J=2$ & 1.5863 & & $J=3$ & 1.4904 \\
$J=4$ & 2.8153 & & $J=5$ & 1.5101 \\
$J=6$ & 3.2420 & & $J=7$ & 0.6163 \\
\botrule
\end{tabular*}
\end{center}

Note that, for the $(j^2)$ configuration, the states with even $J$ have isospin
$T=1$ and those with odd $J$ have isospin $T=0$.

It should be noted that the above energies differ from those of M.B.Z.~\cite{
mbz64} because in 1963 the $T=0$ states were not well known. The M.B.Z. values
for $T=1$ were ($0, 1.509, 2.998$, and $3.400$), while for $T=0$ they were
($1.035, 2.248, 1.958$, and $0.617$).

\section{Calculation of the wave functions}

For previous works on the $f_{7/2}$ shell, we advise the reader to consult
Refs.~\cite{bmz63,mbz64,gf63,g65,g66,kbo78}.

We present results for up to four protons and for any number of neutrons. With
a charge independent interaction, such as the one we use here, this covers all 
cases because the nuclei not included are either mirror nuclei or cross 
conjugate nuclei. A mirror nucleus is one in which we change protons into 
neutrons and neutrons into protons. Assuming charge symmetry, the spectra of
mirrors are identical. This result holds for multishell wave functions. 
For an explanation of cross conjugate nuclei, see Subsect.~\ref{sec:ccn}.
% In the
% single $j$ shell, if we change protons into neutron holes and neutrons into 
% proton holes, we obtain a nucleus thas has the same spectrum as the original
% nucleus and the wave function coefficients $D^I(J_P,J_N)$ are the same---this
% is cross conjugate symmetry. As an example, the cross conjugate of $^{47}$Sc
% is $^{49}$Ti.

If the neutron number is four or less, we use the coefficients of fractional
parentage of the Bayman--Lande method previously described. For neutron number
greater than four, we switch to neutron holes. Operationally, we perform the 
same calculation as before, but for the proton--neutron interaction we 
substitute the proton--neutron-hole interaction, which can be obtained by a 
Pandya relation
\begin{equation}
E(j j^{-1} J) = \text{constant} - \sum_{J'} (2J'+1) 
\begin{Bmatrix} j & j & J' \\ j & j & J \end{Bmatrix} E(j j J') ,
\end{equation}
or, what is equivalent, from the calculated spectrum of $^{48}$Sc. We do not 
change the proton--proton interaction and we note that the spectrum of two
neutron holes is the same as that of two neutrons. The values of the 
proton--neutron-hole matrix elements, with the lowest state energy ($J=6^+$) 
set to zero for $J=0,1,\cdots,7$ are, respectively, (in MeV): 7.03763, 2.39516,
0.32755, 0.39826, 0.12027, 0.08821, 0.00000, 1.18812.

\subsection{Cross conjugate nuclei} \label{sec:ccn}

In the single-$j$-shell model, consider a nucleus with $\mathcal{P}$ valence
protons and $\mathcal{N}$ valence neutrons. We define the cross conjugate 
nucleus as one with $\mathcal{N}$ proton holes and $\mathcal{P}$ neutron holes,
i.e., ($2j+1-\mathcal{N}$) protons and ($2j+1-\mathcal{P}$) neutrons. In the
single-$j$-shell model, the spectra of these two nuclei are identical. If the
wave function of the original nucleus is
\begin{equation}
\Psi = \sum D^I (J_P, J_N) \left[ (j^\mathcal{P}_\pi)^{J_P} (j^\mathcal{N}_\nu
)^{J_N} \right]^I 
\end{equation}
and the wave function of the cross conjugate nucleus is
\begin{equation}
\Phi = \sum C^I (J_N, J_P) \left[ (j^{2j+1-\mathcal{N}}_\pi)^{J_N} 
(j^{2j+1-\mathcal{P}}_\nu)^{J_P} \right]^I ,
\end{equation}
then, for $N\ne 4, Z \ne 4$,
\begin{equation}
C^I(J_N,J_P)= D^I(J_P,J_N) (-1)^{J_P+J_N-I} .
\end{equation}
For $N=4, Z \ne 4$ or $N \ne 4, Z=4$,
\begin{equation}
C^I(J_N,J_P) = D^I (J_P,J_N) (-1)^{J_P+J_N-I} (-1)^{v_4/2} ,
\end{equation}
where $v_4$ is the seniority of the four-particle system.

\noindent For $N=Z=4$, if we consider 4 protons and 4 neutron holes, we get
\begin{equation}
\Psi = \sum_{J_P, J_N} D^I (J_P,J_N) (-1)^{v_P/2} \left[ (j^4)^{J_P v_P}
(j^{-4})^{J_N v_N} \right]^I .
\end{equation}

The wave function of a mirror nucleus is
\begin{eqnarray}
\Psi & = & \sum D^I(J_P,J_N) \left[ (j^\mathcal{P}_\nu)^{J_P} (j^\mathcal{N}
_\pi )^{J_N} \right]^I \nonumber \\
 & = & \sum D^I (J_P,J_N) (-1)^{J_P+J_N-I} \left[ (j^\mathcal{N}_\pi )^{J_N}
(j^\mathcal{P}_\nu)^{J_P} \right]^I . \label{mirror}
\end{eqnarray}

Here is a list of nuclei and their cross conjugates:
\begin{center}
\begin{tabular*}{.7\textwidth}{@{\extracolsep{\fill}}ccccc}
\toprule
Nucleus & Cross conjugate & & Nucleus & Cross conjugate \\
\cline{1-2} \cline{4-5} 
$^{43}$Sc & $^{53}$Fe & & $^{44}$V & $^{52}$Co \\ 
$^{44}$Sc & $^{52}$Mn & & $^{45}$V & $^{51}$Fe \\
$^{45}$Sc & $^{51}$Cr & & $^{46}$V & $^{50}$Mn \\
$^{46}$Sc & $^{50}$V & & $^{47}$V & $^{49}$Cr \\
$^{47}$Sc & $^{49}$Ti & & $^{48}$V & $^{48}$V \\
$^{48}$Sc & $^{48}$Sc \\
$^{43}$Ti & $^{53}$Co & & $^{45}$Cr & $^{51}$Co \\
$^{44}$Ti & $^{52}$Fe & & $^{46}$Cr & $^{50}$Fe \\
$^{45}$Ti & $^{51}$Mn & & $^{47}$Cr & $^{49}$Mn \\
$^{46}$Ti & $^{50}$Cr & & $^{48}$Cr & $^{48}$Cr \\
$^{47}$Ti & $^{49}$V & & \\
$^{48}$Ti & $^{48}$Ti \\
\botrule
\end{tabular*}
\end{center}

\section{Format of tables}

The wave functions for a system of $p$ protons and $n$ neutrons with total
angular momentum $I$ are represented as column vectors
\begin{equation}
\Psi^{I \alpha} = \sum_{J_P J_N} D^{I \alpha} (J_P, J_N) \left[ (j^p)^{J_P} 
(j^n)^{J_N} \right]^I .
\end{equation}
We list the excitation energies at the top; below we list $J_P, J_N$ and the
corresponding $D^{I \alpha}$. 

Note that, as we go down the column, we find the orthonormality condition
\begin{equation}
\sum{J_P, J_N} D^{I \alpha} (J_P, J_N) D^{I \alpha'} (J_P, J_N) = 
\delta_{\alpha \alpha'} .
\end{equation}
We also have along the row the completeness condition
\begin{equation}
\sum_\alpha D^{I \alpha} (J_P,J_N) D^{I \alpha} (J'_P, J'_N) = 
\delta_{J_P J'_P} \delta_{J_N J'_N} .
\end{equation}

Unless specified otherwise, the isospin of the states is the lowest possible
one: $T=|N-Z|/2$. The higher isospin states are specified. One can determine
the isospin of the states by adding to the original interaction a term $a+b
t(1)\cdot t(2)$. This will not change the wave functions or the relative 
excitation energies for states of a given isospin, but it will cause a $T(T+1)$
splitting of states of different isospins.

As an example, we consider the two $I=7/2$ states in $^{43}$Sc with excitation
energies 0.00000 and 4.14201~MeV. The wave functions are, respectively

\begin{center}
\begin{tabular*}{.7\textwidth}{@{\extracolsep{\fill}}ll}
$E=0.00000$ & $0.78776 [j (j^2)^0]^{7/2} + 0.56165 [j (j^2)^2]^{7/2}$ \\
 & $\text{} +0.22082 [j (j^2)^4]^{7/2} + 0.12340 [j (j^2)^6]^{7/2}$
\end{tabular*}
\end{center}
\begin{center}
\begin{tabular*}{.7\textwidth}{@{\extracolsep{\fill}}ll}
$E=4.14201$ & $-0.50000 [j (j^2)^0]^{7/2} + 0.37268 [j (j^2)^2]^{7/2}$ \\
 & $\text{} +0.50000 [j (j^2)^4]^{7/2} + 0.60093 [j (j^2)^6]^{7/2}$
\end{tabular*}
\end{center}

The 4.142~MeV state is assigned isospin $T=3/2$. Note in this case that the 
$D(j,J_N)$'s are identical with the $I=7/2$ coefficients of fractional 
parentage for 3 identical particles in Table~\ref{tab:cfp23}.

As another example, consider the $I=1/2$ lowest energy state ($E^*=
2.68294$~MeV) of $^{47}$Cr. This nucleus is not in the tables, but we can get
its wave function from the mirror nucleus $^{47}$V using the prescription of
Eq.~(\ref{mirror}). Thus, the wave function is
\begin{eqnarray}
\Psi & = & - 0.24622 [2, 3/2] - 0.04793 [2*,3/2] + 0.62848 [2, 5/2] 
 + 0.25132 [2*,5/2] \nonumber \\
 & & \text{} + 0.12809 [4,7/2] + 0.65385 [4*,7/2] - 0.01776 [4, 9/2] 
 + 0.09999 [4*, 9/2] \\
 & & \text{}  - 0.00413 [5, 9/2] - 0.01591 [5,11/2] - 0.15530 [6,11/2] 
 - 0.01738 [8,11/2] , \nonumber 
\end{eqnarray}
where the asterisk indicates seniority 4. Note that there is more ($J_P=2$, 
$v=2$) admixture in the wave function than there is ($J_P=2$, $v=4$). On the 
other hand, there is more ($J_P=4$, $v=4$) than ($J_P=4$, $v=2$) admixture.

\section{Justification}

We feel that a good justification for publishing the single-$j$-shell wave 
functions is the educational value they contain.

There are some striking observations that the reader can ponder:

\begin{enumerate}
\item The even $J$ states in $^{42}$Sc ($J=0,2,4$, and 6) have isospin 1, and
the odd $J$ states ($J=1,3,5$, and 7) have isospin 0.

\item In the even--even Ti isotopes, there is one $I=0$ state with isospin
$T=T_{\text{min}}+2$, where $T_{\text{min}}=|N-Z|/2$. There are no $I=0$ states
with $T=T_{\text{min}}+1$.

\item The $D(J_P,J_N)$'s for $I=j$ in $^{43}$Sc are the same as those of $I=0$
in $^{44}$Ti; and the excitation energies in $^{43}$Sc ($I=j$) are half of 
those in $^{44}$Ti ($I=0$).

\item Except for some overall phases, the $D(J_P,J_N)$'s for the Scandium 
states with isospin $|N-Z|/2 +1$ are the coefficients of fractional parentage
$[j^n J_N j |\} j^{n+1} I v ]$.

\item In $^{44}$Ti the isospin $T=0$ and $T=2$ states are such that $D(J_P,J_N)
=D(J_N,J_P) (-1)^{J_P+J_N-I}$, while for $T=1$, $D(J_P,J_N)=D(J_N,J_P)
(-1)^{J_P+J_N+1-I}$.

\item In self-conjugate $^{48}$Ti, a given state has either $D(J_P,J_N)=
D(J_N,J_P)$ or $D(J_P,J_N)=-D(J_N,J_P)$. Thus, we can assign a signature 
quantum number to the states. Note that the lowest two $6^+$ states have 
opposite signatures; both have isospin $T=2$ and are nearly degenerate.

\item In $^{46}$Ti $I=0$, the column vector for the unique $T=3$ state has 
$D(J_P,J_N)$ identical to the $D(j,J_N)$ for the $T=5/2$ state of $^{45}$Sc
with $I=j$. This implies an equality between two-particle cfp's and 
one-particle cfp's.

\item Note that, in $^{48}$Cr, $(-1)^{(v_P+v_N)/2}$ is a good quantum number, 
where $v_P$ and $v_N$ are the seniorities of the protons and neutrons. This 
leads to striking visual patterns in the wave functions.

\item Because of the small difference in excitation energy of the $19/2^-$ and
$15/2^-$ states in $^{43}$Sc ($^{43}$Ti), the $19/2^-$ state is isomeric. 
Likewise the $12^+$ and $10^+$ states in $^{44}$Ti. However, in the cross
conjugate nucleus $^{52}$Fe, the $12^+$ comes below the $10^+$ and so it has
a much bigger lifetime.

\item Note that the spectrum of $^{48}$Sc (particle--hole) is almost the 
inverted spectrum of $^{42}$Sc. With a $Q\cdot Q$ interaction, the spectrum of
$^{48}$Sc would be precisely the upside down spectrum of $^{42}$Sc.

\item Note that the single-$j$-shell model predicts the near fivefold
degeneracy for the ground state of $^{46}$Sc. The spins involved are $I=2,3,4,
5$, and 6. In $^{45}$Ti the calculation leads to a near degeneracy of the
states with $I=5/2^-$ and $7/2^-$.

\item There is an interesting experimental systematic for the Sc isotopes. The
ground state spins for $^{42}$Sc, $^{44}$Sc, $^{46}$Sc, and $^{48}$Sc are,
respectively, $I=0,2,4$, and 6.

\item There is a dominance of $J=0$ and $J=2$ couplings in the lowest-lying
states of the even--even nuclei. This lends some support to theoretical models
which truncate to these two couplings, such as the interacting boson model.

\item In the low-lying states of systems with 4 neutrons or 4 protons, there is
more admixture of $I=4$, $v=4$ than there is of $I=4$, $v=2$. This is in part 
due to the fact that in $^{44}$Ca the $I=4$, $v=4$ state is slightly lower in 
energy than $I=4$, $v=2$. If $I=4$ is important, then seniority truncation will
not work.

\item We note that the idea of taking matrix elements from experiment is 
present in the book of de~Shalit and Talmi~\cite{st63}. For the upper half of 
the $f_{7/2}$ shell, one might use the spectrum of $^{54}$Co (hole--hole) 
rather than that of $^{42}$Sc.

\item Note that, for the even--even and odd--even nuclei considered here, the
states of higher isospin are at high excitation energies. These states have a
simpler structure than their lower isospin neighbours. Assuming the neutron
has isospin $+1/2$ and the proton $-1/2$, these higher isospin states in a 
nucleus $(N,Z)$ can be obtained by applying the isospin lowering operator to
a state in the nucleus $(N+1,Z-1)$.
\end{enumerate}

\clearpage
\renewcommand{\headrule}{}

\vspace*{5cm}
\begin{table}[h]
\caption{}
\vspace{1cm}
{\large Wave functions of the Scandium isotopes: $^{43-48}$Sc.}
\end{table}
\clearpage \newpage
\lhead{\sf SCANDIUM 43} \pagestyle{fancy}
\begin{Verbatim}
SCANDIUM 43


               I=   0.5

              4.31596
  Jp   Jn 
  3.5  4.0    1.00000


               I=   1.5

              2.88884  5.51324
  Jp   Jn 		T=3/2
  3.5  2.0    0.88641 -0.46291
  3.5  4.0    0.46291  0.88641


               I=   2.5

              3.44955  3.93618  4.47038
  Jp   Jn 			 T=3/2
  3.5  2.0    0.44269 -0.43922  0.78174
  3.5  4.0    0.42803  0.86959  0.24618
  3.5  6.0    0.78792 -0.22563 -0.57296


               I=   3.5

              0.00000  2.79305  4.14201  4.39375
  Jp   Jn 			 T=3/2
  3.5  0.0    0.78776 -0.35240 -0.50000  0.07248
  3.5  2.0    0.56165  0.73700  0.37268  0.04988
  3.5  4.0    0.22082 -0.37028  0.50000 -0.75109
  3.5  6.0    0.12340 -0.44219  0.60093  0.65431


               I=   4.5

              1.68041  4.10973  6.23970
  Jp   Jn 			 T=3/2
  3.5  2.0    0.92524 -0.20187 -0.32121
  3.5  4.0    0.37901  0.45497  0.80582
  3.5  6.0    0.01653  0.86732 -0.49747






               I=   5.5

              2.33560  4.41290  5.95152
  Jp   Jn 			 T=3/2
  3.5  2.0    0.81474 -0.24169 -0.52705
  3.5  4.0    0.50641  0.73931  0.44381
  3.5  6.0    0.28238 -0.62850  0.72474


               I=   6.5

              3.50013  4.95078
  Jp   Jn 
  3.5  4.0    0.98921 -0.14647
  3.5  6.0    0.14647  0.98921


               I=   7.5

              3.51123  7.29248
  Jp   Jn 		T=3/2
  3.5  4.0    0.87905 -0.47673
  3.5  6.0    0.47673  0.87905


               I=   8.5

              4.29802
  Jp   Jn 
  3.5  6.0    1.00000


               I=   9.5

              3.64486
  Jp   Jn 
  3.5  6.0    1.00000

\end{Verbatim}
\clearpage \newpage \lhead{\sf SCANDIUM 44}
\begin{Verbatim}
SCANDIUM 44


               I=   0.0

              3.04737
  Jp   Jn 	T=2
  3.5  3.5    1.00000


               I=   1.0

              0.43199  2.35020  4.48954
  Jp   Jn 
  3.5  2.5    0.86910  0.47980 -0.12024
  3.5  3.5   -0.49126  0.86562 -0.09673
  3.5  4.5    0.05768  0.14314  0.98802


               I=   2.0

              0.00000  2.57532  4.03107  4.63367  6.64525
  Jp   Jn 				   T=2	    T=2
  3.5  1.5    0.13715  0.12709  0.76581 -0.25355 -0.56061
  3.5  2.5   -0.65731  0.50957  0.09103  0.52440 -0.15811
  3.5  3.5    0.73552  0.35451  0.00059  0.57735 -0.00000
  3.5  4.5    0.00023  0.45193  0.38545 -0.27817  0.75485
  3.5  5.5    0.09024  0.62791 -0.50663 -0.50000 -0.30151


               I=   3.0

              0.76421  2.21307  3.12626  3.73904  4.40561
  Jp   Jn 
  3.5  1.5   -0.00156 -0.70281  0.69317 -0.05310 -0.15081
  3.5  2.5   -0.67682  0.47215  0.44711 -0.34468 -0.01691
  3.5  3.5    0.68487  0.48603  0.52442 -0.00567  0.14031
  3.5  4.5   -0.18588 -0.11549  0.10615  0.29923  0.92266
  3.5  5.5   -0.19576  0.18323  0.18254  0.88815 -0.32554











               I=   4.0

              0.71318  2.36326  3.69983  4.82963  5.64017  5.86267
  Jp   Jn 					    T=2	     T=2
  3.5  1.5   -0.40909  0.73685  0.29095  0.20769 -0.17593  0.36187
  3.5  2.5   -0.19846 -0.52492  0.48605  0.01841 -0.65828  0.12309
  3.5  3.5    0.78013  0.03425  0.20467  0.12246 -0.00000  0.57735
  3.5  4.5   -0.32027 -0.35126 -0.48690  0.49993  0.12839  0.52016
  3.5  5.5   -0.28009 -0.22461  0.45254 -0.38879  0.64550  0.31382
  3.5  7.5    0.06026 -0.08065  0.44198  0.73515  0.32026 -0.38925


               I=   5.0

              1.27558  2.05876  3.28296  4.07960  4.63932  6.97396
  Jp   Jn 						     T=2
  3.5  1.5   -0.20317  0.24749  0.79069  0.25383 -0.24051  0.38730
  3.5  2.5   -0.06539  0.90135 -0.03516 -0.26005  0.05533 -0.33371
  3.5  3.5    0.90144  0.17460  0.07483  0.38296  0.06833  0.00000
  3.5  4.5   -0.12578  0.29107 -0.60516  0.34085 -0.33204  0.55391
  3.5  5.5   -0.34788  0.01193 -0.04167  0.77706  0.24949 -0.45939
  3.5  7.5   -0.07082  0.10470 -0.00405 -0.03601  0.87289  0.46987


               I=   6.0

              0.38139  2.21591  3.29471  4.81545  6.28937
  Jp   Jn 					    T=2
  3.5  2.5    0.66438 -0.67732 -0.20671 -0.01719  0.23837
  3.5  3.5    0.66663  0.41410  0.13496 -0.18051 -0.57735
  3.5  4.5    0.17858  0.00171  0.84011  0.43695  0.26718
  3.5  5.5   -0.26488 -0.53596  0.45500 -0.50373 -0.42640
  3.5  7.5   -0.11028 -0.28724 -0.16199  0.72281 -0.59720


               I=   7.0

              1.27227  3.34540  3.68544  5.63639
  Jp   Jn 
  3.5  3.5    0.94213  0.33317 -0.00587  0.03684
  3.5  4.5    0.16788 -0.44518  0.87009 -0.12873
  3.5  5.5   -0.27725  0.81923  0.43577 -0.24921
  3.5  7.5   -0.08569  0.14030  0.23023  0.95915




               I=   8.0

              3.09709  4.12331  8.42688
  Jp   Jn 			  T=2
  3.5  4.5   -0.44265  0.81499 -0.37398
  3.5  5.5    0.84294  0.23597 -0.48349
  3.5  7.5    0.30579  0.52926  0.79144


               I=   9.0

              3.38962  5.44401
  Jp   Jn 
  3.5  5.5    0.98002 -0.19892
  3.5  7.5    0.19892  0.98002


               I=  10.0

              4.79327
  Jp   Jn 
  3.5  7.5    1.00000


               I=  11.0

              4.62848
  Jp   Jn 
  3.5  7.5    1.00000

\end{Verbatim}
\clearpage \newpage \lhead{\sf SCANDIUM 45}
\begin{Verbatim}
SCANDIUM 45


               I=   0.5

              2.09586  4.36606
  Jp   Jn 
  3.5  4.0    0.53421  0.84535
  3.5  4.0*   0.84535 -0.53421


               I=   1.5

              1.65462  2.30236  4.47103  5.36666  7.97281
  Jp   Jn 					   T=5/2
  3.5  2.0    0.88048 -0.15515 -0.30148 -0.21332 -0.25355
  3.5  2.0*   0.35580 -0.24894  0.22229  0.66913  0.56061
  3.5  4.0    0.13926  0.78373 -0.35433 -0.07153  0.48550
  3.5  4.0*   0.27808  0.45820  0.80911 -0.04843 -0.23604
  3.5  5.0   -0.03796  0.29964 -0.28193  0.70661 -0.57446


               I=   2.5

              1.57419  2.93769  4.12260  4.27667  5.31154  6.92995
  Jp   Jn 						    T=5/2
  3.5  2.0    0.76212 -0.39725  0.09354 -0.25828 -0.05081 -0.42817
  3.5  2.0*  -0.01460  0.05628  0.80686  0.44723 -0.35905 -0.12910
  3.5  4.0    0.17008  0.72507 -0.13101 -0.32918 -0.54922 -0.13484
  3.5  4.0*   0.55540  0.25943  0.18695  0.06176  0.25584  0.72111
  3.5  5.0   -0.08114  0.42382  0.34181 -0.23365  0.69212 -0.40414
  3.5  6.0    0.27382  0.25764 -0.41386  0.75267  0.14957 -0.31382


               I=   3.5

              0.00000  2.35344  2.92781  4.33180  4.56413  5.53888  6.60158
  Jp   Jn 							     T=5/2
  3.5  0.0    0.85532  0.36497  0.06105  0.02627  0.17471 -0.01693  0.31623
  3.5  2.0   -0.49579  0.61239 -0.08273  0.09897  0.43226  0.09496  0.40825
  3.5  2.0*   0.03569 -0.20613  0.16501  0.91187  0.09008  0.29899 -0.00000
  3.5  4.0   -0.09578 -0.45436  0.49845 -0.07135  0.22189 -0.42615  0.54772
  3.5  4.0*  -0.10753  0.44830  0.59345  0.09801 -0.64295 -0.11084  0.00000
  3.5  5.0    0.00683 -0.10262  0.45530 -0.37829  0.15468  0.78428 -0.00000
  3.5  6.0    0.02372  0.17706  0.39276  0.01463  0.53663 -0.30382 -0.65828





               I=   4.5

              1.67399  2.68338  3.15738  4.11137  5.27535  6.36697  8.69927
  Jp   Jn 							     T=5/2
  3.5  2.0    0.78666  0.19766 -0.52109 -0.12955  0.14214 -0.05126  0.17593
  3.5  2.0*  -0.25182  0.31593  0.18117 -0.56403  0.40131 -0.31120  0.47741
  3.5  4.0   -0.23048  0.66529 -0.28022 -0.29488 -0.37159  0.07691 -0.44137
  3.5  4.0*   0.50025  0.11684  0.72738 -0.21946 -0.36819 -0.10690 -0.10894
  3.5  5.0   -0.06949  0.39853  0.00125  0.58113 -0.38216 -0.28739  0.51962
  3.5  6.0    0.09757  0.48479  0.28400  0.42423  0.61979  0.19092 -0.27247
  3.5  8.0   -0.00622  0.10495  0.08593 -0.11088 -0.12213  0.87415  0.43613


               I=   5.5

              1.55135  2.81325  3.38660  4.45836  4.59270  6.11404  8.41109
  Jp   Jn 							     T=5/2
  3.5  2.0    0.73843  0.55988 -0.04800 -0.01168  0.21109 -0.10450 -0.28868
  3.5  2.0*   0.11830  0.17299  0.23615  0.79875 -0.44258  0.19009  0.17408
  3.5  4.0   -0.21854  0.54189  0.56017 -0.43874 -0.28776 -0.10206  0.24309
  3.5  4.0*   0.57097 -0.33576 -0.12038 -0.34051 -0.40485  0.13010  0.50000
  3.5  5.0    0.09961 -0.05581  0.39236  0.05108  0.66725  0.48002  0.39340
  3.5  6.0    0.23702 -0.49689  0.67725  0.00407 -0.03909 -0.28130 -0.39696
  3.5  8.0   -0.02937  0.01481  0.03198 -0.22539 -0.25711  0.78489 -0.51472


               I=   6.5

              2.77109  3.46356  4.42594  4.81835  6.90363
  Jp   Jn 
  3.5  4.0   -0.22592  0.91273 -0.01826  0.31814 -0.11973
  3.5  4.0*   0.81369  0.06759 -0.45235  0.35876  0.00215
  3.5  5.0    0.07620 -0.23470  0.65018  0.69233 -0.19251
  3.5  6.0    0.52020  0.29861  0.55423 -0.53600 -0.21390
  3.5  8.0    0.10223  0.13453  0.25522  0.05889  0.95018


               I=   7.5

              2.42857  3.69144  3.89569  5.30439  9.75205
  Jp   Jn 					   T=5/2
  3.5  4.0    0.51787  0.74594  0.03922  0.32507 -0.26112
  3.5  4.0*   0.60799 -0.25701 -0.70964 -0.12064  0.21483
  3.5  5.0   -0.31919 -0.20071 -0.43229  0.74130 -0.34847
  3.5  6.0    0.48065 -0.56663  0.45548  0.09283 -0.48147
  3.5  8.0    0.17104 -0.12714  0.31707  0.56712  0.72967


               I=   8.5

              3.62314  4.98068  6.80088
  Jp   Jn 
  3.5  5.0   -0.36855  0.92490 -0.09339
  3.5  6.0    0.89183  0.32344 -0.31626
  3.5  8.0    0.26230  0.19985  0.94406


               I=   9.5

              4.11046  6.04853
  Jp   Jn 
  3.5  6.0    0.98758 -0.15714
  3.5  8.0    0.15714  0.98758


               I=  10.5

              6.29914
  Jp   Jn 
  3.5  8.0    1.00000


               I=  11.5

              6.34935
  Jp   Jn 
  3.5  8.0    1.00000

\end{Verbatim}
\clearpage \newpage \lhead{\sf SCANDIUM 46}
\begin{Verbatim}
SCANDIUM 46


               I=   0.0

              4.94873
  Jp   Jn 	T=3
  3.5  3.5    1.00000


               I=   1.0

              1.20729  2.36113  3.10960
  Jp   Jn 
  3.5  2.5    0.71482  0.30589  0.62886
  3.5  3.5    0.69301 -0.43021 -0.57849
  3.5  4.5    0.09359  0.84932 -0.51951


               I=   2.0

              0.00000  1.24517  2.14770  3.64329  6.53504
  Jp   Jn 					    T=3
  3.5  1.5    0.08977  0.87809  0.17016 -0.32592 -0.29277
  3.5  2.5    0.47424  0.32101  0.15247  0.53116  0.60553
  3.5  3.5    0.87364 -0.25115 -0.13797 -0.20863 -0.33333
  3.5  4.5    0.01349 -0.20179  0.89911  0.21801 -0.32121
  3.5  5.5   -0.06004  0.14867 -0.34694  0.72152 -0.57735


               I=   3.0

              0.09872  1.67065  2.28180  2.61361  4.31683
  Jp   Jn 
  3.5  1.5    0.15682  0.00539 -0.58270  0.76193  0.23518
  3.5  2.5    0.66326  0.58221 -0.29184 -0.36841  0.01487
  3.5  3.5    0.72635 -0.54553  0.34068  0.16864 -0.17411
  3.5  4.5    0.04743 -0.44700 -0.22432 -0.41268  0.75982
  3.5  5.5    0.07527  0.40445  0.63946  0.29155  0.58038











               I=   4.0

              0.06636  0.61393  1.95829  2.78529  4.52571  7.76403
  Jp   Jn 						     T=3
  3.5  1.5    0.20931  0.14247 -0.12839 -0.76573  0.39807  0.41786
  3.5  2.5    0.69417 -0.67607  0.05817  0.17490  0.08297  0.14213
  3.5  3.5    0.68544  0.61810 -0.08037  0.04632 -0.16855 -0.33333
  3.5  4.5    0.01028  0.27563 -0.42976  0.57184  0.22687  0.60063
  3.5  5.5    0.05596  0.25218  0.88301  0.13944  0.05296  0.36237
  3.5  7.5   -0.03538  0.03191  0.09625  0.18565  0.86717 -0.44947


               I=   5.0

              0.28047  0.77029  2.01461  2.60727  3.91006  5.20082
  Jp   Jn 
  3.5  1.5   -0.28673 -0.16416  0.74430  0.57569  0.07009 -0.02260
  3.5  2.5    0.54199  0.74936  0.35348  0.00994  0.10828 -0.08904
  3.5  3.5   -0.75046  0.55725  0.03942 -0.23706 -0.25516 -0.05846
  3.5  4.5   -0.13183 -0.16318  0.26915 -0.55569  0.67035 -0.35409
  3.5  5.5   -0.20832  0.27207 -0.42703  0.45143  0.67595  0.19882
  3.5  7.5   -0.00811 -0.01794  0.25438 -0.31575  0.10943  0.90732


               I=   6.0

              0.03554  1.90106  2.83419  3.64485  8.19073
  Jp   Jn 					    T=3
  3.5  2.5   -0.42369  0.80164 -0.13247  0.29077  0.27524
  3.5  3.5    0.86241  0.22717 -0.21618  0.21633  0.33333
  3.5  4.5   -0.07775 -0.31418  0.54509  0.70919  0.30852
  3.5  5.5    0.26480  0.45370  0.68874 -0.08515 -0.49237
  3.5  7.5    0.02391 -0.03473 -0.40527  0.59871 -0.68958


               I=   7.0

              1.07484  2.46047  3.37742  4.90167
  Jp   Jn 
  3.5  3.5    0.95913 -0.19378  0.15177 -0.13961
  3.5  4.5   -0.11665  0.27484  0.94087 -0.16006
  3.5  5.5    0.25776  0.83637 -0.13324  0.46508
  3.5  7.5   -0.00541 -0.43290  0.27199  0.85941




               I=   8.0

              1.90941  3.35025  3.67269
  Jp   Jn 
  3.5  4.5    0.54717  0.82417  0.14607
  3.5  5.5    0.73801 -0.39272 -0.54873
  3.5  7.5   -0.39489  0.40806 -0.82314


               I=   9.0

              2.83527  4.02713
  Jp   Jn 
  3.5  5.5    0.91524  0.40290
  3.5  7.5   -0.40290  0.91524


               I=  10.0

              4.26919
  Jp   Jn 
  3.5  7.5    1.00000


               I=  11.0

              4.67440
  Jp   Jn 
  3.5  7.5    1.00000

\end{Verbatim}
\clearpage \newpage \lhead{\sf SCANDIUM 47}
\begin{Verbatim}
SCANDIUM 47


               I=   0.5

              2.88876
  Jp   Jn 
  3.5  4.0    1.00000


               I=   1.5

              1.31504  2.65087
  Jp   Jn 
  3.5  2.0    0.99825  0.05912
  3.5  4.0   -0.05912  0.99825


               I=   2.5

              1.43194  2.97711  5.19793
  Jp   Jn 
  3.5  2.0    0.98139  0.03522  0.18874
  3.5  4.0   -0.13342  0.83200  0.53849
  3.5  6.0   -0.13807 -0.55365  0.82122


               I=   3.5

              0.00000  2.30919  4.07704  8.82950
  Jp   Jn 				  T=7/2
  3.5  0.0    0.92859  0.25110 -0.19739  0.18898
  3.5  2.0   -0.37019  0.67513 -0.47810  0.42258
  3.5  4.0   -0.02520 -0.68848 -0.45159  0.56695
  3.5  6.0   -0.00699  0.08451  0.72700  0.68139


               I=   4.5

              1.96638  3.03249  4.97236
  Jp   Jn 
  3.5  2.0    0.83441  0.47712  0.27590
  3.5  4.0   -0.54382  0.63139  0.55281
  3.5  6.0    0.08956 -0.61131  0.78631






               I=   5.5

              1.44487  2.98044  3.42366
  Jp   Jn 
  3.5  2.0    0.92605 -0.07778  0.36929
  3.5  4.0   -0.37577 -0.28088  0.88312
  3.5  6.0   -0.03503  0.95659  0.28934


               I=   6.5

              2.87525  3.67730
  Jp   Jn 
  3.5  4.0   -0.51200  0.85898
  3.5  6.0    0.85898  0.51200


               I=   7.5

              2.56966  3.75863
  Jp   Jn 
  3.5  4.0   -0.64027  0.76815
  3.5  6.0    0.76815  0.64027


               I=   8.5

              3.63633
  Jp   Jn 
  3.5  6.0    1.00000


               I=   9.5

              4.44012
  Jp   Jn 
  3.5  6.0    1.00000

\end{Verbatim}
\clearpage \newpage \lhead{\sf SCANDIUM 48}
\begin{Verbatim}
SCANDIUM 48


               I=   0.0

              7.03763
  Jp   Jn 	T=4
  3.5  3.5    1.00000


               I=   1.0

              2.39516
  Jp   Jn 
  3.5  3.5    1.00000


               I=   2.0

              0.32755
  Jp   Jn 
  3.5  3.5    1.00000


               I=   3.0

              0.39826
  Jp   Jn 
  3.5  3.5    1.00000


               I=   4.0

              0.12027
  Jp   Jn 
  3.5  3.5    1.00000


               I=   5.0

              0.08821
  Jp   Jn 
  3.5  3.5    1.00000


               I=   6.0			               I=   7.0

              0.00000                                 1.18812
  Jp   Jn				  Jp   Jn
  3.5  3.5    1.00000			  3.5  3.5    1.00000
\end{Verbatim}

\clearpage \newpage \lhead{}
\vspace*{5cm}
\begin{table}[h]
\caption{}
\vspace{1cm}
{\large Wave functions of the Titanium isotopes: $^{44-48}$Ti.}
\end{table}
\clearpage \newpage \lhead{\sf TITANIUM 44}
\begin{Verbatim}
TITANIUM 44


               I=   0.0

              0.00000  5.58610  8.28402  8.78750
  Jp   Jn 			  T=2
  0.0  0.0    0.78776 -0.35240 -0.50000  0.07248
  2.0  2.0    0.56165  0.73700  0.37268  0.04988
  4.0  4.0    0.22082 -0.37028  0.50000 -0.75109
  6.0  6.0    0.12340 -0.44219  0.60093  0.65431


               I=   1.0

              5.66864  7.58685  9.72619
  Jp   Jn 	T=1	 T=1	  T=1
  2.0  2.0    0.90585 -0.31554 -0.28262
  4.0  4.0    0.40144  0.42645  0.81055
  6.0  6.0    0.13524  0.84769 -0.51296


               I=   2.0

              1.16313  4.95650  5.23665  7.81197  7.82336  7.96963  9.26771  9.87032
  Jp   Jn 			  T=1	   T=1			      T=1      T=2
  0.0  2.0    0.60986  0.15576 -0.63698  0.30702 -0.12984 -0.06012 -0.00051 -0.28868
  2.0  0.0    0.60986  0.15576  0.63698 -0.30700 -0.12988 -0.06012  0.00051 -0.28868
  2.0  2.0   -0.35385  0.80997  0.00000  0.00000 -0.00621 -0.18411  0.00000 -0.26937
  2.0  4.0    0.24163  0.18510 -0.29324 -0.60808  0.40660  0.19472  0.21043  0.38686
  4.0  2.0    0.24163  0.18510  0.29324  0.60803  0.40668  0.19473 -0.21043  0.38686
  4.0  4.0   -0.05913  0.37942  0.00000 -0.00000 -0.02378  0.64143 -0.00000 -0.04307
  4.0  6.0    0.06128  0.18100 -0.09092 -0.18977  0.02494 -0.47024 -0.67507  0.31382
  6.0  4.0    0.06128  0.18100  0.09093  0.18977  0.02497 -0.47024  0.67507  0.31382
  6.0  6.0    0.05625  0.13198  0.00000  0.00000 -0.79606  0.17134 -0.00000  0.51247

             11.88190
  Jp   Jn 	T=2
  0.0  2.0    0.00000
  2.0  0.0    0.00000
  2.0  2.0   -0.33503
  2.0  4.0   -0.23328
  4.0  2.0   -0.23328
  4.0  4.0    0.66234
  4.0  6.0    0.37848
  6.0  4.0    0.37848
  6.0  6.0   -0.23177



               I=   3.0

              5.78578  6.00086  7.44971  8.36291  8.69411  8.97569  9.64226
  Jp   Jn 		 T=1	  T=1	   T=1		     T=1      T=1
  2.0  2.0    0.00000 -0.76123  0.51299  0.09331  0.00000 -0.34305 -0.17597
  2.0  4.0   -0.69657  0.40439  0.48148 -0.21572  0.12161 -0.13282 -0.20118
  4.0  2.0    0.69657  0.40439  0.48148 -0.21572 -0.12161 -0.13282 -0.20118
  4.0  4.0    0.00000 -0.22508  0.15490 -0.74542 -0.00000  0.24226  0.55768
  4.0  6.0   -0.12161  0.12114  0.28729  0.39789 -0.69657  0.01550  0.49420
  6.0  4.0    0.12161  0.12114  0.28729  0.39789  0.69657  0.01550  0.49420
  6.0  6.0    0.00000 -0.11599  0.29005  0.16106  0.00000  0.88762 -0.29769


               I=   4.0

              2.79000  5.00025  5.94982  6.70828  7.59990  8.14760  8.28333  8.93648
  Jp   Jn 			  T=1		    T=1			       T=1
  0.0  4.0    0.49966  0.37777 -0.67561 -0.12804  0.02966  0.07765 -0.04324  0.17725
  2.0  2.0    0.64525 -0.49795 -0.00000  0.31754  0.00000 -0.04234  0.16462  0.00000
  2.0  4.0   -0.15474  0.35405 -0.00888  0.47359 -0.64851 -0.01813  0.08450 -0.08587
  2.0  6.0    0.12725  0.15024 -0.19151  0.24698  0.04451 -0.39069 -0.21151 -0.66974
  4.0  0.0    0.49966  0.37777  0.67561 -0.12804 -0.02966  0.07765 -0.04324 -0.17725
  4.0  2.0   -0.15474  0.35405  0.00888  0.47359  0.64851 -0.01813  0.08450  0.08587
  4.0  4.0    0.00551 -0.26231  0.00000  0.44369 -0.00000  0.50902 -0.58566  0.00000
  4.0  6.0    0.04164  0.21965 -0.08240 -0.03490 -0.27672  0.44526  0.04112  0.11250
  6.0  2.0    0.12725  0.15024  0.19151  0.24698 -0.04451 -0.39069 -0.21151  0.66974
  6.0  4.0    0.04164  0.21965  0.08240 -0.03490  0.27672  0.44526  0.04112 -0.11250
  6.0  6.0    0.02328 -0.07401 -0.00000  0.31067  0.00000  0.15681  0.72044 -0.00000

             10.06628 10.87682 11.09932
  Jp   Jn 	T=1	 T=2	  T=2
  0.0  4.0    0.10605 -0.00000 -0.28868
  2.0  2.0   -0.00000 -0.35047  0.28835
  2.0  4.0    0.26830 -0.34339 -0.03210
  2.0  6.0   -0.11310  0.38490  0.23391
  4.0  0.0   -0.10605 -0.00000 -0.28868
  4.0  2.0   -0.26830 -0.34339 -0.03210
  4.0  4.0    0.00000  0.19660 -0.30586
  4.0  6.0   -0.63559  0.10050  0.48862
  6.0  2.0    0.11310  0.38490  0.23391
  6.0  4.0    0.63559  0.10050  0.48862
  6.0  6.0    0.00000  0.53497 -0.26009





               I=   5.0

              5.87126  6.51223  7.02262  7.29541  8.51961  9.31625  9.87596 12.21061
  Jp   Jn 		 T=1		   T=1	    T=1	     T=1      T=1      T=2
  2.0  4.0    0.51771  0.54822  0.35485  0.36348 -0.06808 -0.11928 -0.22017 -0.32567
  2.0  6.0   -0.47871 -0.41374  0.43177  0.47585 -0.18316  0.10044 -0.24239 -0.29054
  4.0  2.0   -0.51771  0.54822 -0.35485  0.36348 -0.06808 -0.11928 -0.22017  0.32567
  4.0  4.0    0.00000 -0.06692 -0.00000  0.14917  0.94486  0.09872 -0.26601  0.00000
  4.0  6.0    0.05305 -0.06616  0.43320  0.32232  0.12390 -0.33783  0.51211  0.55635
  6.0  2.0    0.47871 -0.41374 -0.43177  0.47585 -0.18316  0.10044 -0.24239  0.29054
  6.0  4.0   -0.05305 -0.06616 -0.43320  0.32232  0.12390 -0.33783  0.51211 -0.55635
  6.0  6.0   -0.00000  0.20814  0.00000  0.22990  0.01301  0.84461  0.43621  0.00000


               I=   6.0

              4.06180  5.16735  5.61803  6.88378  7.45256  7.55033  8.53136  9.10583
  Jp   Jn 			  T=1		    T=1		      T=1
  0.0  6.0    0.57535 -0.17867 -0.57732 -0.20536 -0.35862 -0.06041  0.11688 -0.08884
  2.0  4.0    0.34102  0.56158 -0.06649  0.16725 -0.19852  0.03110 -0.63688  0.03887
  2.0  6.0    0.21098 -0.36974 -0.39319  0.41914  0.49401 -0.10222 -0.00055  0.17788
  4.0  2.0    0.34102  0.56158  0.06649  0.16725  0.19852  0.03110  0.63688  0.03887
  4.0  4.0    0.12553 -0.15979 -0.00000 -0.37045 -0.00000  0.80029 -0.00000  0.12537
  4.0  6.0   -0.00325 -0.03252 -0.08767  0.39779  0.29653  0.35259 -0.28411 -0.41176
  6.0  0.0    0.57535 -0.17868  0.57732 -0.20536  0.35862 -0.06041 -0.11688 -0.08884
  6.0  2.0    0.21098 -0.36974  0.39319  0.41914 -0.49401 -0.10223  0.00055  0.17788
  6.0  4.0   -0.00325 -0.03252  0.08767  0.39779 -0.29654  0.35259  0.28411 -0.41176
  6.0  6.0   -0.02366  0.06581 -0.00000  0.23377 -0.00000  0.28417 -0.00000  0.75040

             10.05210 11.52602
  Jp   Jn 	T=1	 T=2
  0.0  6.0    0.15632 -0.28868
  2.0  4.0   -0.22482  0.19462
  2.0  6.0   -0.31836  0.31782
  4.0  2.0    0.22482  0.19462
  4.0  4.0   -0.00000  0.40656
  4.0  6.0    0.56891 -0.21641
  6.0  0.0   -0.15632 -0.28868
  6.0  2.0    0.31836  0.31782
  6.0  4.0   -0.56891 -0.21641
  6.0  6.0    0.00000 -0.54461






               I=   7.0

              6.04279  6.50892  8.37435  8.58205  8.92209 10.87304
  Jp   Jn 		 T=1		   T=1	    T=1      T=1
  2.0  6.0    0.69409  0.51559  0.13503 -0.46604 -0.09218  0.09211
  4.0  4.0   -0.00000  0.26647  0.00000  0.04871  0.89814 -0.34636
  4.0  6.0    0.13503  0.41287 -0.69409  0.45551 -0.24373 -0.25030
  6.0  2.0   -0.69409  0.51559 -0.13503 -0.46603 -0.09218  0.09211
  6.0  4.0   -0.13503  0.41287  0.69410  0.45550 -0.24373 -0.25030
  6.0  6.0   -0.00000  0.23750  0.00000  0.38507  0.23989  0.85894


               I=   8.0

              6.08420  7.34501  8.24456  8.33374  9.35996 13.66353
  Jp   Jn 				   T=1	    T=1      T=2
  2.0  6.0    0.57798  0.12605 -0.26586 -0.59103  0.38818  0.28172
  4.0  4.0    0.49795 -0.69675  0.34922  0.00000 -0.00000 -0.38030
  4.0  6.0    0.20430  0.49062  0.28803 -0.38818 -0.59103 -0.36687
  6.0  2.0    0.57798  0.12605 -0.26586  0.59103 -0.38818  0.28172
  6.0  4.0    0.20430  0.49062  0.28802  0.38818  0.59103 -0.36687
  6.0  6.0    0.02083  0.03670  0.75549  0.00000 -0.00000  0.65380


               I=   9.0

              7.98380  8.62627 10.68065
  Jp   Jn 		 T=1	  T=1
  4.0  6.0   -0.70711  0.57109 -0.41696
  6.0  4.0    0.70711  0.57109 -0.41696
  6.0  6.0   -0.00000  0.58968  0.80764


               I=  10.0

              7.38394  8.90568 10.02992
  Jp   Jn 			  T=1
  4.0  6.0    0.70089 -0.09358  0.70711
  6.0  4.0    0.70089 -0.09358 -0.70711
  6.0  6.0    0.13234  0.99120 -0.00000


               I=  11.0				               I=  12.0

              9.86512				              7.70224	
  Jp   Jn 	T=1				  Jp   Jn
  6.0  6.0    1.00000				  6.0  6.0    1.00000
\end{Verbatim}
\clearpage \newpage \lhead{\sf TITANIUM 45}
\begin{Verbatim}
TITANIUM 45


               I=   0.5

              3.28051  5.24733  6.30068  7.33854  8.57087
  Jp   Jn 			 T=3/2		   T=3/2
  2.0  1.5    0.82783  0.38940  0.34204  0.03406  0.21193
  2.0  2.5   -0.14899 -0.26630  0.84251  0.29097 -0.33526
  4.0  3.5   -0.53633  0.69419  0.25183  0.09067  0.39850
  4.0  4.5   -0.04178 -0.50712  0.24926 -0.34517  0.74820
  6.0  5.5    0.05565 -0.19590 -0.21825  0.88702  0.35225


               I=   1.5

              1.51774  2.23019  4.65627  5.51352  5.85943  6.50717  7.34413  7.81724
  Jp   Jn 					   T=3/2    T=3/2
  0.0  1.5    0.68666  0.23938  0.13021  0.26683 -0.20092  0.54201  0.03732 -0.08159
  2.0  1.5   -0.36221  0.02112 -0.14807  0.62336 -0.51816 -0.15523 -0.07529 -0.17193
  2.0  2.5    0.04365  0.70143 -0.41339 -0.24054 -0.34471 -0.12617 -0.14352  0.10350
  2.0  3.5   -0.54173  0.47531  0.30135  0.10081  0.22904  0.42867  0.19997  0.10139
  4.0  2.5    0.01075  0.04229  0.71560 -0.25748 -0.48860 -0.26009 -0.11088 -0.11531
  4.0  3.5    0.23352  0.42826  0.25530  0.25333  0.42184 -0.59143  0.13250 -0.00841
  4.0  4.5   -0.09233  0.11602 -0.16114 -0.46289 -0.05556  0.03495  0.45676 -0.52985
  4.0  5.5    0.18120 -0.13966 -0.13768  0.04685 -0.15339 -0.20820  0.59711  0.09970
  6.0  4.5   -0.00450 -0.02524  0.09207 -0.34758 -0.04644  0.05552 -0.30469  0.26757
  6.0  5.5   -0.07674 -0.07240  0.23854  0.01749 -0.24618  0.10918  0.46716  0.26251
  6.0  7.5    0.00245  0.01701 -0.10887 -0.06441 -0.11746 -0.07378  0.16205  0.70586

              8.67584  9.57147 12.17762
  Jp   Jn      T=3/2	T=3/2	 T=5/2
  0.0  1.5   -0.12341 -0.02499  0.15811
  2.0  1.5   -0.13246  0.08656  0.32404
  2.0  2.5    0.20505  0.07988 -0.25071
  2.0  3.5   -0.25637  0.07742 -0.14638
  4.0  2.5   -0.15831 -0.17475 -0.17928
  4.0  3.5    0.05174  0.07802  0.28031
  4.0  4.5   -0.11273 -0.15676  0.45477
  4.0  5.5   -0.42485  0.41168 -0.37839
  6.0  4.5   -0.21337  0.66082  0.46710
  6.0  5.5    0.71886  0.18473  0.14527
  6.0  7.5   -0.27609 -0.52728  0.29277


               I=   2.5

              0.00000  2.77134  3.74818  4.84813  5.10273  5.77900  6.00704  7.14250
  Jp   Jn 						    T=3/2	      T=3/2
  0.0  2.5    0.70897 -0.21780 -0.14007 -0.09740  0.11137 -0.56182  0.15000 -0.01101
  2.0  1.5   -0.06222  0.11221  0.53951 -0.10474  0.47968  0.09791  0.00243  0.08931
  2.0  2.5    0.40158  0.68953  0.00011 -0.25521 -0.03641  0.07582 -0.24787  0.15159
  2.0  3.5    0.51036 -0.18081  0.31198  0.15592 -0.21076  0.51913  0.03299  0.20448
  2.0  4.5    0.15661 -0.26677  0.26362  0.08486  0.28085  0.01421 -0.53581 -0.59038
  4.0  1.5   -0.01909 -0.06804 -0.16613 -0.17273 -0.58264  0.04302 -0.35876 -0.24616
  4.0  2.5   -0.01923  0.37538  0.34511 -0.18875 -0.20432 -0.10845  0.37487 -0.51274
  4.0  3.5    0.13730  0.37159 -0.42082  0.50164  0.26673  0.19644  0.11571 -0.33638
  4.0  4.5    0.05447  0.12263  0.09082  0.06311 -0.13026  0.08457 -0.32139  0.16566
  4.0  5.5   -0.13470  0.13838 -0.11588 -0.21958  0.25342 -0.24725 -0.35117  0.18533
  6.0  3.5   -0.06961  0.16639  0.40783  0.54160 -0.30253 -0.51472 -0.07980  0.10883
  6.0  4.5    0.01512  0.00708  0.00263 -0.25099 -0.08287  0.05526  0.04360 -0.26115
  6.0  5.5    0.00779 -0.10681  0.10225 -0.34228 -0.00248  0.01859  0.27994 -0.00543
  6.0  7.5   -0.03747  0.06097 -0.01511 -0.20027  0.03054 -0.10251 -0.17889  0.00742


              7.39634  8.26436  8.32741  8.48148  9.51635 11.13476
  Jp   Jn 			 T=3/2	  T=3/2	   T=3/2    T=5/2
  0.0  2.5    0.10437  0.00475  0.13878 -0.10486 -0.06309  0.15811
  2.0  1.5    0.25493 -0.08787  0.54285  0.06456 -0.14272  0.20471
  2.0  2.5   -0.18439  0.00033  0.11109  0.01893  0.13831 -0.37528
  2.0  3.5    0.02330 -0.10666 -0.26107  0.28571  0.07469  0.24721
  2.0  4.5   -0.08041  0.09727 -0.14478 -0.01450  0.10638 -0.25394
  4.0  1.5    0.37115  0.00302  0.44555  0.21988  0.02991  0.14639
  4.0  2.5    0.03770  0.15719 -0.22170 -0.12208  0.21770  0.33086
  4.0  3.5    0.27799  0.00111 -0.00657  0.20681 -0.22784  0.07785
  4.0  4.5    0.31522  0.20060 -0.22664 -0.64708 -0.43214  0.13595
  4.0  5.5    0.34225 -0.27560 -0.45733  0.24564  0.26025  0.29621
  6.0  3.5    0.02979 -0.15057 -0.03607  0.23464 -0.14281 -0.18119
  6.0  4.5   -0.17318 -0.77022 -0.08988 -0.07226 -0.46802 -0.03354
  6.0  5.5    0.48409  0.21427 -0.24322  0.28311 -0.27588 -0.53526
  6.0  7.5   -0.43004  0.41071 -0.08855  0.42088 -0.51905  0.32098


               I=   3.5

              0.09268  2.36384  3.23023  3.99595  4.20481  5.09641  5.38973  5.74186
  Jp   Jn 					   T=3/2
  0.0  3.5    0.76556 -0.16197  0.16230  0.18315 -0.34918 -0.15423  0.09174 -0.16179
  2.0  1.5   -0.11989  0.33525 -0.31915  0.52918 -0.02021  0.11497  0.23166 -0.16504
  2.0  2.5   -0.47498 -0.42707  0.41093  0.25259 -0.22534 -0.20589 -0.09861 -0.22744
  2.0  3.5   -0.02688  0.63488  0.40023  0.09747 -0.49398 -0.12983 -0.00000  0.09474
  2.0  4.5    0.16554 -0.01863 -0.09902  0.08736 -0.06860  0.56842 -0.25242 -0.42585
  2.0  5.5   -0.25586  0.26714  0.07685 -0.15388  0.04283 -0.06692  0.26001 -0.53538
  4.0  1.5    0.10701 -0.04850  0.22617 -0.11252  0.16708 -0.06043  0.26665  0.22229
  4.0  2.5   -0.11556 -0.21304 -0.21832  0.55055 -0.07419 -0.02762  0.25687  0.31781
  4.0  3.5   -0.12814 -0.31974  0.17847 -0.15465 -0.39433  0.43130  0.38601  0.02842
  4.0  4.5   -0.11474  0.06540  0.01801 -0.13937 -0.13755  0.01777 -0.27349  0.31000
  4.0  5.5    0.03572 -0.14311  0.03049  0.00905  0.15878 -0.42368 -0.05919 -0.30892
  4.0  7.5   -0.07627  0.07488  0.00030 -0.08428  0.00616  0.10048  0.09075 -0.04240
  6.0  2.5    0.02184  0.04453  0.44351  0.43015  0.31969  0.23980 -0.41731  0.07116
  6.0  3.5   -0.08904 -0.08277 -0.44173  0.01743 -0.43085 -0.23557 -0.34660 -0.03753
  6.0  4.5    0.03502 -0.03860  0.04134  0.07754  0.10093  0.07737  0.29841 -0.04270
  6.0  5.5    0.07779  0.07463  0.03916  0.14353  0.19268 -0.25004  0.20200  0.03698
  6.0  7.5    0.05635 -0.08923  0.00725  0.05980  0.08998 -0.11967 -0.01105 -0.25234

              6.55825  6.83905  7.13262  7.93600  8.53661  8.71674  8.76894  9.74369
  Jp   Jn      T=3/2		 T=3/2		   T=3/2	     T=3/2    T=3/2
  0.0  3.5    0.14900 -0.03257  0.02493  0.12553  0.01072  0.05096  0.07132  0.00691
  2.0  1.5    0.13990 -0.21043 -0.45323  0.06688  0.32109  0.01841  0.01585  0.09310
  2.0  2.5   -0.29917 -0.13282 -0.09842  0.03421  0.06969  0.01054  0.17044  0.08289
  2.0  3.5   -0.17763  0.10507  0.05758 -0.16190 -0.03866 -0.04821 -0.15061  0.04991
  2.0  4.5   -0.40323  0.14711  0.22306 -0.13125  0.28561 -0.13024  0.06197 -0.12432
  2.0  5.5    0.31339 -0.03033  0.45609  0.24633 -0.04084 -0.10650  0.02672 -0.22321
  4.0  1.5   -0.03277 -0.45050  0.28936 -0.03862  0.61104 -0.17771  0.09121  0.16882
  4.0  2.5   -0.02498  0.20080  0.51167 -0.11839 -0.03821 -0.03880 -0.27132 -0.13909
  4.0  3.5    0.36318  0.09735 -0.21005 -0.02530  0.04436  0.08607 -0.03327 -0.19398
  4.0  4.5    0.08343  0.41034 -0.01670  0.48113  0.52075  0.03612 -0.06990 -0.08388
  4.0  5.5    0.25167  0.41767 -0.17142 -0.39852  0.30926 -0.25114 -0.24353  0.00498
  4.0  7.5    0.13100  0.24258  0.20256 -0.38412  0.12456  0.58813  0.25103  0.47627
  6.0  2.5    0.46811 -0.02466  0.09505  0.06073 -0.09352  0.01538  0.10260  0.01745
  6.0  3.5    0.26254 -0.13713  0.21511  0.08664  0.01979 -0.05585  0.33869  0.15153
  6.0  4.5   -0.11548  0.40081 -0.01376  0.38446 -0.17547 -0.36951  0.19137  0.57066
  6.0  5.5   -0.19165  0.25456 -0.05352  0.08658  0.07033  0.27745  0.56848 -0.47818
  6.0  7.5   -0.13111 -0.06427  0.06697  0.39241  0.05649  0.54653 -0.49305  0.14035



             10.80639
  Jp   Jn      T=5/2
  0.0  3.5    0.31623
  2.0  1.5    0.10351
  2.0  2.5    0.21409
  2.0  3.5   -0.23570
  2.0  4.5   -0.11356
  2.0  5.5    0.20412
  4.0  1.5   -0.19821
  4.0  2.5    0.06742
  4.0  3.5   -0.31623
  4.0  4.5    0.28490
  4.0  5.5   -0.17189
  4.0  7.5    0.21320
  6.0  2.5   -0.15691
  6.0  3.5   -0.38006
  6.0  4.5   -0.17588
  6.0  5.5   -0.28069
  6.0  7.5   -0.39312


               I=   4.5

              1.52729  3.13473  3.62684  4.69325  4.91705  5.84718  5.87880  6.13113
  Jp   Jn							     T=3/2 
  0.0  4.5   -0.41185  0.42753 -0.44374  0.05972  0.15835 -0.07695 -0.27568 -0.19550
  2.0  2.5   -0.46274  0.26905  0.36600  0.02389 -0.40936 -0.10365  0.37808  0.16500
  2.0  3.5    0.65207  0.44732  0.10333  0.22503 -0.07961 -0.04242  0.19373  0.08945
  2.0  4.5    0.11617 -0.12429  0.09640  0.56250 -0.01923 -0.10647 -0.21046 -0.34115
  2.0  5.5    0.19820 -0.14728  0.10126 -0.37891  0.32572 -0.31265  0.39325 -0.30998
  4.0  1.5   -0.00271 -0.06141 -0.26933  0.09642  0.35833 -0.47858  0.00285  0.58827
  4.0  2.5   -0.31558 -0.28765  0.29959  0.46020  0.23888 -0.17286  0.25051 -0.05012
  4.0  3.5   -0.12083  0.52644  0.00126  0.07466  0.35703  0.10011  0.33565 -0.07319
  4.0  4.5    0.05058  0.11520  0.08677 -0.01635  0.09620  0.19467  0.09811  0.20526
  4.0  5.5   -0.10137 -0.12769 -0.08299 -0.36106 -0.02059  0.26931  0.11923  0.01151
  4.0  7.5    0.00206  0.00187 -0.06357 -0.09060  0.25371 -0.12317  0.10971 -0.32603
  6.0  1.5    0.02832  0.12961  0.01182  0.03079 -0.15153 -0.02817 -0.06888 -0.33993
  6.0  2.5    0.01928 -0.27137 -0.39479  0.31071  0.03918  0.51869  0.40778  0.08253
  6.0  3.5   -0.06483 -0.00901  0.52998 -0.10438  0.41753  0.29268 -0.32027  0.08568
  6.0  4.5   -0.07272 -0.05395 -0.12831 -0.02125 -0.00646  0.02833  0.23214 -0.27705
  6.0  5.5   -0.01997 -0.12312 -0.06733 -0.09933 -0.31716 -0.32749  0.07477  0.02466
  6.0  7.5   -0.01331 -0.07596 -0.01720 -0.03214 -0.09006 -0.11989  0.01423 -0.06727

              6.88819  7.05271  7.17102  7.36219  8.19481  8.31618  9.48016 10.57178
  Jp   Jn      T=3/2			  T=3/2		    T=3/2    T=3/2    T=3/2
  0.0  4.5   -0.50472  0.01803  0.00223 -0.05859 -0.05087  0.01063  0.02600  0.12263
  2.0  2.5   -0.11091 -0.01670 -0.00193 -0.33408  0.06645 -0.22973 -0.04094  0.09337
  2.0  3.5   -0.41742  0.02995  0.14567  0.20801 -0.03642 -0.05424 -0.05030  0.05461
  2.0  4.5    0.13288  0.17318 -0.23042 -0.16812  0.06793 -0.51273 -0.13037  0.19332
  2.0  5.5   -0.16413 -0.05472  0.02826 -0.36062 -0.00148 -0.02301  0.12599  0.26191
  4.0  1.5    0.11515  0.15500 -0.00350 -0.03021  0.11165 -0.15483  0.09033  0.21295
  4.0  2.5   -0.19996  0.13731  0.21617  0.33197 -0.13430  0.33283 -0.08841 -0.02082
  4.0  3.5    0.49981 -0.02808 -0.00098  0.22652  0.11020 -0.15614  0.13327 -0.16137
  4.0  4.5   -0.05506  0.49871 -0.66991 -0.16159 -0.19711  0.30206 -0.06685  0.01370
  4.0  5.5   -0.07862  0.51108  0.23037  0.35048  0.00168 -0.39877 -0.28256  0.25263
  4.0  7.5   -0.10315  0.15830 -0.08635 -0.10516  0.17973 -0.06287 -0.36734 -0.56778
  6.0  1.5    0.19341  0.47029  0.20588 -0.04906  0.31568  0.35915  0.39495  0.24897
  6.0  2.5   -0.19446 -0.04364  0.04428 -0.23537  0.24374 -0.06701  0.25275 -0.06623
  6.0  3.5   -0.27363 -0.08899 -0.09864  0.07559  0.27018 -0.18940  0.31804  0.03201
  6.0  4.5    0.01935 -0.31844 -0.44325  0.39079 -0.18869 -0.02012  0.06855  0.39582
  6.0  5.5   -0.19216  0.04204 -0.34238  0.37776  0.52673 -0.05116  0.19870 -0.22849
  6.0  7.5   -0.06179  0.22958  0.03243  0.04482 -0.57284 -0.30830  0.58657 -0.36442
             12.90408
  Jp   Jn      T=5/2
  0.0  4.5   -0.15811
  2.0  2.5    0.19670
  2.0  3.5    0.10157
  2.0  4.5   -0.14865
  2.0  5.5   -0.29129
  4.0  1.5    0.28762
  4.0  2.5   -0.10531
  4.0  3.5   -0.25482
  4.0  4.5   -0.12679
  4.0  5.5   -0.05378
  4.0  7.5    0.49202
  6.0  1.5    0.29542
  6.0  2.5    0.02598
  6.0  3.5    0.15731
  6.0  4.5    0.44164
  6.0  5.5   -0.29264
  6.0  7.5    0.07660


               I=   5.5

              1.57089  3.59524  4.19629  4.68521  5.21520  5.75616  6.24598  6.43567
  Jp   Jn 						    T=3/2
  0.0  5.5   -0.55147  0.51844 -0.14792 -0.03957 -0.09059 -0.46525 -0.02796  0.03501
  2.0  3.5    0.68386  0.25986 -0.19292  0.01699  0.04868 -0.14231  0.03634  0.20625
  2.0  4.5   -0.19544 -0.03098 -0.29908 -0.17160  0.50748 -0.13022  0.06772  0.28548
  2.0  5.5    0.17202  0.05416 -0.57196  0.42392 -0.19491 -0.28775 -0.09299 -0.27574
  2.0  7.5   -0.15060  0.14659  0.11649 -0.02514 -0.10047 -0.07420  0.00615 -0.33791
  4.0  1.5   -0.12986 -0.10235 -0.00583 -0.08209 -0.57043  0.04698  0.47603  0.27148
  4.0  2.5   -0.30632 -0.20000 -0.29420  0.46475  0.25845  0.35076  0.12263  0.02338
  4.0  3.5    0.02599  0.56833  0.36844  0.43841  0.08096  0.30994  0.04268  0.04388
  4.0  4.5    0.07332  0.08052 -0.06314  0.15060 -0.21562 -0.05905  0.48831  0.02658
  4.0  5.5   -0.03661  0.14181  0.11117  0.09003 -0.14352 -0.01056 -0.26205  0.55973
  4.0  7.5    0.03290  0.01391 -0.05277  0.13392  0.02458 -0.05423  0.00335 -0.05296
  6.0  1.5    0.06925  0.10298  0.10265  0.04727  0.43059 -0.18571  0.59137  0.07358
  6.0  2.5   -0.09608 -0.33591  0.00172  0.38613 -0.14230 -0.05778 -0.00910  0.28479
  6.0  3.5   -0.02588  0.31784 -0.50198 -0.29251 -0.09750  0.56264 -0.02543  0.15533
  6.0  4.5   -0.02665  0.04783  0.07358  0.29332  0.03027 -0.03559 -0.25489  0.14782
  6.0  5.5    0.04485 -0.10322  0.01998 -0.04070  0.01758 -0.26366 -0.13548  0.33436
  6.0  7.5    0.03398 -0.02543 -0.03072  0.01097  0.03139 -0.07086 -0.01613  0.21737

              6.73644  7.01806  7.59141  7.83152  8.66317  8.79751  9.00581 10.31885
  Jp   Jn 		T=3/2	 T=3/2		   T=3/2    T=3/2	      T=3/2
  0.0  5.5   -0.01673  0.21604  0.15339 -0.11319  0.13542 -0.14894 -0.07311  0.15107
  2.0  3.5   -0.12451  0.49166  0.18431  0.05996  0.14825 -0.05749  0.02937  0.10999
  2.0  4.5    0.26857  0.17723 -0.42008  0.21656 -0.22750  0.06299  0.07931 -0.14746
  2.0  5.5    0.12273 -0.32124 -0.22836 -0.06528  0.11569  0.15962  0.11801 -0.16338
  2.0  7.5   -0.33613  0.15008  0.24771  0.35365 -0.22939  0.38790  0.25402 -0.33678
  4.0  1.5   -0.10716  0.08079 -0.35318  0.04159  0.21950  0.17960  0.13737  0.21560
  4.0  2.5   -0.29134  0.07510  0.23293  0.11783  0.29170 -0.25361  0.00940  0.07866
  4.0  3.5    0.05838  0.06368 -0.40022  0.11038  0.08662 -0.00345  0.04218 -0.18221
  4.0  4.5    0.08794 -0.11345  0.14283  0.08679 -0.55838 -0.49465 -0.17413 -0.18602
  4.0  5.5    0.28417 -0.38906  0.35820  0.13664  0.12405  0.09214 -0.05155 -0.11748
  4.0  7.5    0.00695 -0.09435 -0.00707  0.61409 -0.12957  0.28077 -0.45610  0.49748
  6.0  1.5   -0.08861 -0.30829  0.20979 -0.27989  0.09268  0.39431  0.00606  0.02015
  6.0  2.5    0.18028  0.44835  0.16947 -0.11400 -0.15801  0.33692 -0.05411 -0.25329
  6.0  3.5   -0.14169 -0.07891  0.09693 -0.12810 -0.19437  0.23431 -0.07171 -0.08637
  6.0  4.5   -0.28996 -0.02609 -0.13095 -0.40517 -0.52137  0.09493  0.09409  0.43523
  6.0  5.5   -0.66854 -0.17432 -0.23741  0.09534  0.05981 -0.10522 -0.26423 -0.35724
  6.0  7.5   -0.06555 -0.17817  0.09470  0.30865 -0.10214 -0.14842  0.74868  0.17998

             12.61590
  Jp   Jn      T=5/2
  0.0  5.5   -0.15811
  2.0  3.5    0.16667
  2.0  4.5    0.26591
  2.0  5.5    0.03170
  2.0  7.5    0.32991
  4.0  1.5    0.21847
  4.0  2.5    0.20945
  4.0  3.5   -0.14035
  4.0  4.5    0.04910
  4.0  5.5    0.36875
  4.0  7.5   -0.18549
  6.0  1.5   -0.08387
  6.0  2.5   -0.37848
  6.0  3.5   -0.22918
  6.0  4.5    0.26714
  6.0  5.5   -0.17447
  6.0  7.5   -0.42185


               I=   6.5

              3.48722  4.11274  5.14902  5.36881  5.86469  6.41664  6.64122  6.97590
  Jp   Jn 								      T=3/2
  2.0  4.5    0.41880  0.25247 -0.46150 -0.46122  0.16259  0.01702 -0.09071  0.08175
  2.0  5.5    0.42110 -0.03493  0.38725 -0.30386  0.31462 -0.01747  0.18565 -0.55399
  2.0  7.5   -0.03747 -0.07339 -0.23893  0.16221  0.24519 -0.64405  0.25333  0.26859
  4.0  2.5    0.27513  0.57546  0.29809 -0.01002 -0.39827  0.13066  0.01855  0.39914
  4.0  3.5   -0.70104  0.29675  0.14663 -0.32803  0.18204  0.09489 -0.09599 -0.10650
  4.0  4.5    0.10906 -0.15147  0.21862  0.12931 -0.32509 -0.29902 -0.17281 -0.22009
  4.0  5.5   -0.04382 -0.00948  0.27337  0.26594  0.12101  0.26319  0.61809  0.05412
  4.0  7.5    0.03400 -0.05649  0.00859 -0.05086 -0.02524 -0.16617  0.11157 -0.02673
  6.0  1.5    0.11804 -0.37208 -0.30169  0.18180  0.05920  0.59681 -0.06633  0.04776
  6.0  2.5    0.19374 -0.12861  0.45293  0.16579  0.43513 -0.04753 -0.40637  0.39943
  6.0  3.5    0.07380  0.55843 -0.20280  0.49376  0.37523  0.01928  0.04712 -0.24522
  6.0  4.5   -0.07087 -0.01807  0.08357 -0.09392  0.39271  0.10958 -0.13844  0.23278
  6.0  5.5   -0.00367 -0.13055  0.04381 -0.39626  0.08584  0.00297  0.40313  0.32136
  6.0  7.5    0.04243 -0.00417 -0.02864  0.02686 -0.08067  0.03741  0.32525  0.11009

              7.42645  7.66837  8.28800  8.63075  9.02316 11.10844
  Jp   Jn 		T=3/2		  T=3/2	   T=3/2    T=3/2
  2.0  4.5    0.06473  0.20776 -0.01413 -0.43926 -0.18655 -0.13496
  2.0  5.5   -0.06334  0.14353  0.08395  0.18142  0.21311  0.16885
  2.0  7.5   -0.07128  0.39530  0.17971  0.08687  0.30474  0.06725
  4.0  2.5   -0.03152  0.24754  0.03427  0.23381  0.22523  0.00123
  4.0  3.5   -0.00384  0.43027  0.13337  0.00861 -0.14997  0.05644
  4.0  4.5    0.58877  0.31486  0.18454 -0.06186 -0.25290 -0.28217
  4.0  5.5   -0.01657  0.15914 -0.06574 -0.50790  0.00398 -0.30946
  4.0  7.5   -0.20265  0.25112 -0.75740  0.32719 -0.34850 -0.21283
  6.0  1.5    0.01449  0.48153  0.16100  0.31433  0.01521 -0.01205
  6.0  2.5   -0.19575  0.04704  0.08864 -0.11169 -0.35471  0.10162
  6.0  3.5    0.15203 -0.14077  0.09518  0.26127 -0.25267 -0.10083
  6.0  4.5    0.62734 -0.08469 -0.39127  0.07082  0.41507 -0.09026
  6.0  5.5    0.14417 -0.28614  0.34108  0.39409 -0.31503 -0.27801
  6.0  7.5    0.34890  0.05569 -0.13115 -0.05605 -0.32688  0.78664





               I=   7.5

              3.23706  4.55610  5.27236  6.13842  6.63338  6.64792  6.77842  7.53075
  Jp   Jn 					   T=3/2
  0.0  7.5    0.37198 -0.61651 -0.23015 -0.11539 -0.47334  0.01182 -0.09469 -0.10413
  2.0  5.5    0.56961 -0.12482  0.56725  0.18378 -0.04938  0.09990  0.20428  0.14101
  2.0  7.5   -0.04950 -0.14045 -0.43241  0.47307 -0.29645 -0.04622 -0.07686  0.32520
  4.0  3.5   -0.65851 -0.25295  0.22754  0.25421 -0.20718  0.07741  0.20152  0.05014
  4.0  4.5    0.09604  0.17842  0.27960  0.52515 -0.03142 -0.08919 -0.50305 -0.06836
  4.0  5.5    0.02845  0.19695 -0.28003  0.19039 -0.12052  0.54060 -0.10002 -0.30526
  4.0  7.5   -0.03884 -0.01677  0.10914  0.23795 -0.08991  0.05488  0.32866  0.44860
  6.0  1.5    0.13503  0.20108 -0.16578 -0.07826  0.00000  0.50234  0.44182  0.13806
  6.0  2.5    0.24125  0.06785 -0.35007  0.48177  0.41489 -0.20359  0.25415 -0.11593
  6.0  3.5   -0.07502 -0.60474 -0.05115  0.05035  0.60559  0.05092  0.03747 -0.05178
  6.0  4.5    0.08262  0.13084 -0.22750 -0.23430  0.09936 -0.20763 -0.22307  0.64421
  6.0  5.5    0.02568  0.14642 -0.06310 -0.00157 -0.25963 -0.57330  0.46885 -0.27276
  6.0  7.5    0.00300  0.02049 -0.09805 -0.01308 -0.05430 -0.10184 -0.03034 -0.19316

              7.89625  8.10051  8.39035  9.50920 13.95686
  Jp   Jn      T=3/2	T=3/2		  T=3/2	   T=5/2
  0.0  7.5    0.10075  0.29634 -0.14252  0.16728 -0.15811
  2.0  5.5    0.23638 -0.20141  0.20302 -0.10536  0.28571
  2.0  7.5   -0.26933 -0.24256  0.21824 -0.37814  0.20825
  4.0  3.5    0.44770  0.26406  0.03187  0.00626  0.15076
  4.0  4.5    0.07972  0.19110 -0.30180 -0.23034 -0.38898
  4.0  5.5    0.39871 -0.40954  0.15166  0.24552 -0.16064
  4.0  7.5   -0.24247 -0.24002 -0.37436  0.50933 -0.30553
  6.0  1.5    0.02509  0.32205 -0.26846 -0.49323 -0.14638
  6.0  2.5    0.12325  0.38387  0.04697  0.29586  0.19656
  6.0  3.5    0.08998 -0.30628 -0.09080 -0.25034 -0.27798
  6.0  4.5    0.58910 -0.02207 -0.02097  0.00359 -0.06055
  6.0  5.5    0.21323 -0.23567  0.05262 -0.21535 -0.36534
  6.0  7.5    0.14302 -0.28337 -0.74153 -0.07262  0.53136


               I=   8.5

              3.72637  5.32155  6.47859  6.50287  7.31953  7.82796  7.86267  9.18549
  Jp   Jn 						    T=3/2	      T=3/2
  2.0  7.5   -0.01498  0.45249 -0.37136  0.48295  0.20039  0.55196 -0.22448 -0.06098
  4.0  4.5    0.06068  0.13988  0.65900  0.55176  0.10289 -0.20199 -0.14580 -0.29603
  4.0  5.5   -0.16983  0.52391  0.45717 -0.34489  0.25442  0.25393  0.28914  0.39330
  4.0  7.5   -0.01014  0.10420 -0.25007  0.05249  0.01396  0.06525  0.74983 -0.44435
  6.0  2.5    0.69763  0.51782 -0.14431 -0.02607 -0.19940 -0.40055  0.04567  0.11804
  6.0  3.5    0.67250 -0.42981  0.12593  0.00149  0.34458  0.42042  0.07864  0.15247
  6.0  4.5    0.10873  0.18214  0.01287 -0.58189  0.17500  0.11058 -0.43187 -0.60193
  6.0  5.5   -0.12641  0.01928 -0.33774  0.03282  0.64838 -0.42475 -0.15320  0.27187
  6.0  7.5    0.01988 -0.06263  0.07317  0.01585  0.52423 -0.23660  0.25347 -0.28827

             11.00569
  Jp   Jn      T=3/2
  2.0  7.5   -0.15796
  4.0  4.5    0.27885
  4.0  5.5    0.03494
  4.0  7.5    0.39950
  6.0  2.5   -0.08630
  6.0  3.5    0.14909
  6.0  4.5    0.15668
  6.0  5.5    0.41269
  6.0  7.5   -0.71549

			
               I=   9.5

              4.22791  5.86444  6.93292  7.70053  8.31527  9.16755 10.25334
  Jp   Jn 					   T=3/2	     T=3/2
  2.0  7.5   -0.18558  0.53650 -0.58041 -0.00014  0.25146  0.08670  0.51971
  4.0  5.5   -0.29884  0.70205  0.30080  0.22239  0.07031 -0.07184 -0.51748
  4.0  7.5   -0.16592 -0.07716 -0.08690 -0.68851  0.47010 -0.46064 -0.22745
  6.0  3.5    0.83807  0.17960  0.01888  0.18276  0.46555 -0.09700 -0.07408
  6.0  4.5    0.19302  0.35066  0.57337 -0.51846 -0.21088  0.11061  0.43072
  6.0  5.5   -0.31852 -0.21733  0.48220  0.36994  0.52522 -0.16551  0.42272
  6.0  7.5   -0.08753 -0.10439  0.05882 -0.19347  0.41686  0.85211 -0.20169


               I=  10.5

              6.34667  7.40568  8.85968 10.50395
  Jp   Jn 				  T=3/2
  4.0  7.5    0.16326  0.65618  0.29703  0.67420
  6.0  4.5   -0.42698  0.71936 -0.17930 -0.51775
  6.0  5.5    0.87573  0.22790 -0.30043 -0.30151
  6.0  7.5    0.15538  0.00286  0.88846 -0.43183


               I=  11.5

              6.53122  8.04457 10.55416
  Jp   Jn 			 T=3/2
  4.0  7.5    0.37886  0.86626  0.32567
  6.0  5.5    0.87299 -0.21773 -0.43644
  6.0  7.5    0.30716 -0.44966  0.83873


               I=  12.5

              8.46781
  Jp   Jn 
  6.0  7.5    1.00000


               I=  13.5

              7.88789
  Jp   Jn 
  6.0  7.5    1.00000

\end{Verbatim}
\clearpage \newpage \lhead{\sf TITANIUM 46}
\VerbatimInput[obeytabs=true,xleftmargin=-1cm,baselinestretch=.8,firstline=4]
{ti46-1.txt}
\clearpage \newpage \lhead{\sf TITANIUM 47}
\VerbatimInput[obeytabs=true,xleftmargin=-1cm,baselinestretch=.76,firstline=4]
{ti47h-1.txt}
\clearpage \newpage \lhead{\sf TITANIUM 48}
\VerbatimInput[obeytabs=true,xleftmargin=-1cm,baselinestretch=.9,firstline=4]
{ti48h-1.txt}

\clearpage \newpage \lhead{}
\vspace*{5cm}
\begin{table}[h]
\caption{}
\vspace{1cm}
{\large Wave functions of the Vanadium isotopes: $^{46-48}$V.}
\end{table}
\clearpage \newpage \lhead{\sf VANADIUM 46}
\VerbatimInput[obeytabs=true,xleftmargin=-1cm,baselinestretch=.75,firstline=4]
{v46-1.txt}
\clearpage \newpage \lhead{\sf VANADIUM 47}
\VerbatimInput[obeytabs=true,xleftmargin=-1cm,baselinestretch=.7,firstline=4]
{v47-1.txt}
\clearpage \newpage \lhead{\sf VANADIUM 48}
\VerbatimInput[obeytabs=true,xleftmargin=-1cm,baselinestretch=.75,firstline=4]
{v48h-1.txt}

\clearpage \newpage \lhead{}
\vspace*{5cm}
\begin{table}[h]
\caption{}
\vspace{1cm}
{\large Wave functions of the Chromium isotopes: $^{48}$Cr.}
\end{table}
\clearpage \newpage \lhead{\sf CHROMIUM 48}
\VerbatimInput[obeytabs=true,xleftmargin=-1cm,baselinestretch=.7,firstline=4]
{cr48-1.txt}

\end{document}